\documentclass[prb,aps,amsmath,amsfonts,superscriptaddress]{revtex4}

\usepackage{graphicx}

\def\Tr{\mbox{Tr}\,}
\def\tr{\mbox{tr}\,}
\newcommand{\R} {\mbox{Re}\,}
\newcommand{\I} {\mbox{Im}\,}

\begin{document}

\title{Factorization of quantum charge 
 transport for non-interacting fermions}

\author{Alexander G.~Abanov}
%\footnote{E-mail:  alexandre.abanov@sunysb.edu}}
\affiliation{Department of Physics and Astronomy,
Stony Brook University,  Stony Brook, NY 11794-3800.}

\author{D.~A.~Ivanov}
\affiliation{Institute of Theoretical Physics,
Ecole Polytechnique F\'ed\'erale de Lausanne (EPFL), 
CH-1015 Lausanne, Switzerland}

\date{April 21, 2009}

\begin{abstract}
We show that the statistics of the charge transfer of non-interacting
fermions through a two-lead contact is generalized binomial, at any
temperature and for any form of the scattering matrix: an arbitrary
charge-transfer process can be decomposed into independent single-particle
events. This result generalizes previous studies of adiabatic pumping at
zero temperature and of transport induced by bias voltage.
\end{abstract}

\maketitle

\tableofcontents

%%%%%%%%%%%%%%%%%%%%%%%%%%%%%%
%%%%%%%%%%%%%%%%%%%%%%%%%%%%%%
\section{Introduction}
%%%%%%%%%%%%%%%%%%%%%%%%%%%%%%
%%%%%%%%%%%%%%%%%%%%%%%%%%%%%%

Noise is usually considered as an unwanted characteristics 
of electronic circuits. Sometimes, however, studying electronic noise gives 
important information about physical systems. 
In particular, noise in small junctions is affected by the fermionic
statistics of electrons and is an essential part of quantum 
mesoscopic transport (see, e.g., Ref.~\onlinecite{2000-BlanterButtiker}). 
Even more information
about the quantum behavior of charge carriers is encoded in the
full probability distribution of the transferred charge: a concept
known as full counting statistics \cite{1993-LevitovLesovik}. 

The full counting statistics of electrons is most studied
for the simplest model of non-interacting fermions in various setups.
In this model, the general expression for the probability distribution
of the transferred charge is given by an elegant determinant formula
of Levitov and Lesovik \cite{1993-LevitovLesovik,1993-IvanovLevitov,%
1996-LevitovLeeLesovik,1997-IvanovLeeLevitov}. 
This formula expresses the generating function for the probability distribution as a
certain functional determinant involving the single-particle scattering matrix.
An exact calculation of this functional determinant may be performed in several
special setups of the charge transfer \cite{1993-LevitovLesovik,1993-IvanovLevitov,%
1996-LevitovLeeLesovik,1997-IvanovLeeLevitov,2001-Levitov,2008-VanevicNazarovBelzig},
and until recently most
of the results on the full counting statistics in this model concentrated 
on studying such exactly solvable cases.

However in the recent years a progress has been made in understanding
general properties of charge transfer encoded in the determinant formula. 
Namely, it has been shown that, in the case of a contact with two external 
leads, the total electronic transfer is given by
a superposition of uncorrelated elementary charge transfers of single
electrons. In each of such single-electron events, one electron has
a certain probability $p$ to pass through the contact (thereby transferring
one quantum of charge) and the probability $1-p$ to reflect (resulting in
no charge transfer). This decomposition into elementary charge-transfer
events (dubbed ``generalized binomial'' statistics \cite{2008-Hassler})
has been shown under various assumptions: at zero temperature for
a charge transfer driven by a time-dependent bias voltage 
\cite{2007-VanevicNazarovBelzig}, at zero temperature for the adiabatic-pumping
problem \cite{2008-AbanovIvanov}, and in the wave-packet formalism
(a finite number of wave packets with a momentum-dependent scattering)
\cite{2008-Hassler}.

The goal of the present paper is to lift the assumptions made in those previous
works and to show that the generalized binomial statistics is universally valid
for any charge-transfer problem involving non-interacting fermions: at any
temperature and for any time- and energy-dependent scatterer. A necessary 
condition remains that the initial state does not involve any
entanglement between the leads or between different charge sectors in
each lead.

It may be therefore practical to describe the statistics for noninteracting
fermions by the distribution $\mu(p)$ of the effective transparencies
of the elementary charge transfers, instead of the probabilities of the
total transfer. To illustrate this construction, we list several known
examples where the distribution $\mu(p)$ is exactly known. Finally, to
illustrate the importance of the non-interacting assumption, we also
present several examples of {\it interacting} systems where the full
counting statistics is {\it not} generalized binomial. In those cases,
we may interpret the effect of interaction as a shift
of the transparencies $p$ to the complex plane.

The paper is organized as follows. First, to fix the notation, we
briefly review the derivation of the Levitov--Lesovik determinant 
formula in Section~\ref{sec:detFCS}. In Section~\ref{sec:FCSzeros} 
we obtain our main result: a decomposition of an arbitrary charge 
transfer process into independent tunneling events. 
In Sections \ref{sec:FCSnonint} and \ref{sec:FCSint}, we illustrate
our construction with several noninteracting and interacting examples,
respectively. Finally, we conclude with Section~\ref{sec:Conclusion},
where we discuss possible generalizations and applications of our results. 
Some technical details are relegated to Appendices
A, B, C.

%%%%%%%%%%%%%%%%%%%%%%%%%%%%%%
%%%%%%%%%%%%%%%%%%%%%%%%%%%%%%
\section{Generating function and determinant formula}
\label{sec:detFCS}
%%%%%%%%%%%%%%%%%%%%%%%%%%%%%%
%%%%%%%%%%%%%%%%%%%%%%%%%%%%%%

In this section we briefly review some known results from the
theory of full counting statistics for non-interacting fermions
(mainly following the method of Ref.~\onlinecite{2002-Klich}). 
We also introduce the notation and specify the assumptions used in 
the subsequent sections.

%%%%%%%%%%%%%%%%%%%%%%%%%%%%%%
\subsection{Generating function for full counting statistics}

A typical full-counting-statistics problem involves determining the probabilities
of a given charge transfer through the junction. In this paper, we will consider
the case of a junction with two leads, and therefore the definitions in this
section will be written for a two-lead junction, even though they admit a
natural generalization to a many-lead situation \cite{1992-Buttiker}. 
In the case of two leads,
the charge transfer is characterized by one number $q$: an integer number of
electrons moved from the left to the right lead (or $-q$ electrons from
the right to the left lead). The ingredients of the problem are:
(i) the initial density matrix of the system $\hat{\rho}_{0}$ and (ii) the unitary
evolution operator $\hat{U}$ (over the time when the measurement is performed).
To formally define the transferred charge $q$, we also need (iii) 
an operator $\hat{Q}$
which counts the particles in one of the leads (say, the right lead).

Then one easily sees that the probabilities of a given charge transfer $q$ can
be found from the generating function
\begin{equation}
	\chi(\lambda)=\Tr\left(\hat{\rho}_{0} \hat{U}^{\dagger}
	e^{i\lambda\hat{Q}}\hat{U}e^{-i\lambda \hat{Q}}\right)\Big/\,  
	\Tr\hat{\rho}_{0} \, ,
 \label{mbFCS}
\end{equation}
where $\lambda$ is an auxiliary variable and the trace is taken in the many-body Fock space
(for technical reasons, it is convenient to explicitly include the normalization of the
density matrix).
For a good definition of charge transfer {\it probabilities}, one needs to
further assume that different charge sectors are not entangled, i.e., that $\hat{\rho}_{0}$ 
commutes with $\hat{Q}$. Under this assumption one finds that thus defined $\chi(\lambda)$
may indeed be interpreted as a generating function
\begin{equation}
	\chi(\lambda) = \sum_{q=-\infty}^{\infty}P_{q}\, e^{i\lambda q}\, ,
 \label{FCS}
\end{equation}
for some probabilities $0\leq P_q\leq 1$. The periodicity of $\chi(\lambda)$ 
(or, equivalently, the quantization of $q$) follows from the fact that the spectrum
of $\hat{Q}$ is integer. It will therefore be convenient for our purposes to treat
$\chi(\lambda)$ as a function of the complex variable 
\begin{equation}
	u=e^{i\lambda}\, .
\end{equation}
Then (\ref{FCS}) becomes the Laurent series of the complex function $\chi(u)$.
The Fourier components of $\chi(\lambda)$ (the coefficients of the Laurent series) 
define the probabilities $P_q$, while the derivatives at $\lambda=0$ give the moments 
of the charge transfer $\langle q^n \rangle$.

In the case of a periodic pumping, it may be useful to define a ``full counting
statistics per period'' \cite{1997-IvanovLeeLevitov} as the extensive part 
of $\chi(\lambda)$:
\begin{equation}
	\chi_0(\lambda)=\exp\left[\lim_{N_p\to\infty} \frac{1}{N_p} \ln \chi(\lambda)\right]\, ,
 \label{chi-extensive}
\end{equation}
where the full counting statistics $\chi(\lambda)$ is collected over $N_p$ periods.
Defined in this way, the characteristic function $\chi_0(\lambda)$ may be used 
for calculating extensive quantities (such as cumulants of the transferred charge), 
but the effects associated with the beginning and ending of the 
measurement \cite{1993-LevitovLesovik,1996-LevitovLeeLesovik} are neglected.

%%%%%%%%%%%%%%%%%%%%%%%%%%%%%%
\subsection{Determinant formula for noninteracting fermions}

In the case of non-interacting fermions, the general expression  (\ref{mbFCS})
may be simplified and rewritten in terms of {\it single-particle} operators: the so-called
Levitov--Lesovik determinant formula \cite{1993-LevitovLesovik,1993-IvanovLevitov,%
1996-LevitovLeeLesovik,1997-IvanovLeeLevitov}. In this section we briefly present
its derivation following the approach of Ref.~\onlinecite{2002-Klich}. 

For non-interacting fermions, the multi-particle operators (acting in the
Fock space) are related to the corresponding single-particle operators
(acting in the Hilbert space of one particle) with the help of
the second-quantized formalism. For notational convenience, we will use
the ``hatted'' notation (as in Eq.~\ref{mbFCS}) for operators acting
in the Fock space, while the ``unhatted'' letters will denote the
corresponding single-particle operators. For quantum-mechanical observables
(such as the charge operator), the correspondence is given by
$\hat{Q}=\psi^{\dagger}Q \psi$ (where $\psi^\dagger$ and $\psi$ are
the fermionic creation and annihilation operators, and the summation
over single-particle states is assumed), while for their exponentials
(such as the evolution operator or the density matrix) the correspondence
is $\hat{U}=\exp[\psi^\dagger (\ln U) \psi]$ (see Appendix~\ref{app:fbil}
for details).

Now we are ready to derive the Levitov--Lesovik determinant formula,
starting from (\ref{mbFCS}) and making the following assumptions:
\vspace{0.4cm}
\\
\textit{1) The operator $\hat{Q}$ is a fermionic bilinear.} 
This is obviously true for the operator of the charge in the right lead. 
However, more generally, one can use the same formalism for 
studying full counting statistics for other physical quantities (e.g., spin).
\vspace{0.2cm}
\\
\textit{2) The evolution operator $\hat{U}$ is an exponential of 
a fermionic bilinear.} 
This is equivalent to requiring that the Hamiltonian is a fermionic bilinear
(see Appendix~\ref{app:fbil}), i.e., that the fermions are non-interacting.
\vspace{0.2cm}
\\
\textit{3) The density matrix $\hat{\rho}_{0}$ is an exponential of 
a fermionic bilinear.} $\hat{\rho}_{0}=e^{-\hat{h}_{0}}$, 
where $\hat{h}_{0}=\psi^{\dagger}h_{0}\psi$ is a Hermitian
fermionic bilinear (note that, if using Eq.~(\ref{mbFCS}), we do not need to 
normalize the density matrix). Usually, one considers a thermal equilibrium 
with $h_0$ given by the physical Hamiltonian divided by the temperature. 
However, our assumption is much
less restrictive. For example, one can take an initial state with different 
temperatures of the leads and/or of different channels or even prepare an 
equilibrium state using a different Hamiltonian.
\vspace{0.2cm}
\\
\textit{4) The operators $\hat{Q}$ and $\hat{\rho}_{0}$ commute.} 
This condition expresses the absence of charge entanglement in the initial state
and is a necessary condition for interpreting $\chi(\lambda)$ as a generating
function for charge-transfer probabilities.
\vspace{0.4cm}

The trace of the product of exponentials of fermionic bilinears in (\ref{mbFCS}) can 
be rewritten as a determinant of an operator acting in the single-particle Hilbert space
(see Ref.~\onlinecite{2002-Klich} and Appendix~\ref{app:fbil} for details):
\begin{equation}
	\chi(\lambda)= \det\Big[1+n_{F}(U^{\dagger}e^{i\lambda Q}Ue^{-i\lambda Q}-1)\Big]\, .
 \label{spFCS}
\end{equation}
Here $U$ is a single-particle evolution operator and
\begin{equation}
	n_{F}=\frac{\rho_{0}}{\rho_{0}+1}=\frac{1}{1+e^{h_{0}}}
 \label{nF}
\end{equation}
is the occupation-number operator.

%%%%%%%%%%%%%%%%%%%%%%%%%%%%%%
\subsection{Regularization of the determinant}

Depending on the dimensionality and on the asymptotic behavior
of the operators involved in (\ref{spFCS}), the determinant may
require a regularization.

If the dimension of the single-particle Hilbert space is finite
(this is the case in the wave-packet formalism of 
Ref.~\onlinecite{2008-Hassler}), all the matrices in 
(\ref{spFCS}) are finite-dimensional, and no regularization is needed.

If the single-particle Hilbert space has an infinite dimension, then
a regularization may be needed or not, depending on the properties
of the Hamiltonians describing the evolution (in the operator $\hat{U}$)
and the initial state (in the operator $\hat{\rho}$). Usually,
one considers a certain dispersion of the particle in the leads (the
same in the evolution and the initial-state Hamiltonians), with the
fermions filling the states up to some Fermi energy. 
The determinant (\ref{spFCS}) is well-defined, if the matrix tends
to unity sufficiently fast at both large positive and negative energies. At large
positive energies, $n_F$ tends to zero, and the matrix under determinant
tends to unity. At large negative energies, the determinant is cut off
either by the Fermi energy or by the energy dependence of $U$: at
very low energy, the particle cannot penetrate through the barrier,
then $U$ commutes with $Q$, and the matrix again tends to unity. So
the determinant (\ref{spFCS}) is well defined if one uses exact
dispersion relation and the energy dependence of the scattering matrix.

The only (but historically very common in literature)
case when the determinant needs a regularization is the approximation
where the spectrum is linearized (so that the Fermi energy does not
serve as a regularizing parameter) and the scattering matrix is assumed
to be energy-independent (``instant scattering'' or ``adiabatic pumping''
limit) \cite{1993-IvanovLevitov,1998-Brouwer,2003-MuzykantskyAdamov}. 
In that case, the matrix in (\ref{spFCS}) is
infinite-dimensional and tends to a non-unity matrix in the negative-energy
asymptotics, therefore the determinant needs to be regularized. The 
regularization may be performed by admitting a weak energy dependence
of $U$, so that the determinant converges, and then letting the
corresponding energy scale tend to infinity. Technically, this
procedure amounts to simply re-expressing the determinant in a
manifestly convergent form where the weak energy dependence
of $U$ becomes inessential.  This approach was used
in Ref.~\onlinecite{2003-MuzykantskyAdamov} and then more 
generally (but also more formally) in 
Ref.~\onlinecite{2007-AvronBachmannGrafKlich}. For completeness,
we review the details of adiabatic approximation in 
Appendix~\ref{app:adiabatic}. 

Therefore, we assume, without loss of generality, that an appropriate
energy dependence is already included in the evolution operator $U$, and
the determinant (\ref{spFCS}) does not need any further regularization.
Moreover, in the future discussion we will be interested in the zeros of 
the determinant, which are regularization independent.

%%%%%%%%%%%%%%%%%%%%%%%%%%%%%%
%%%%%%%%%%%%%%%%%%%%%%%%%%%%%%
\section{Decomposition into single-particle events}
\label{sec:FCSzeros}
%%%%%%%%%%%%%%%%%%%%%%%%%%%%%%
%%%%%%%%%%%%%%%%%%%%%%%%%%%%%%

%%%%%%%%%%%%%%%%%%%%%%%%%%%%%%
\subsection{Zeros of the generating function}

To demonstrate that the full counting statistics for non-interacting
fermions can be decomposed into elementary single-particle transfer
events, we first re-express the determinant (\ref{spFCS}) in terms of the
spectral properties of a single operator, whose eigenvalues give
transmission probabilities of those elementary transfers.

This transformation proceeds as follows. 
Using the fact that the charge operator $Q$ is a projector
($Q^{2}=Q$), we rewrite 
$ e^{i\lambda Q}=1+(e^{i\lambda}-1)Q $ in (\ref{spFCS}), 
in order to obtain, after some simple algebra,
\begin{equation}
	\chi(\lambda) = 
  \det\left(\left[1+(e^{i\lambda}-1)X\right]e^{-i\lambda Q}\right)\, .
 \label{XlambdaFCS}
\end{equation}
Here we have introduced a new operator,
\begin{equation}
	X = (1-n_{F})Q +n_{F} U^{\dagger}Q U\, .
 \label{X}
\end{equation}

One can see that the spectrum of the operator $X$ is
\begin{enumerate}
	\item Real,
	\item Confined to the interval $[0,1]$.
\end{enumerate}

Both properties become obvious if one considers a Hermitian operator 
conjugate to $X$ by a similarity transformation:
\begin{equation}
   \tilde{X}=(n_{F})^{-1/2}X(n_{F})^{1/2} =(1-n_{F})Q +
      (n_{F})^{1/2}U^{\dagger}Q U (n_{F})^{1/2}
\label{tX}
\end{equation}
(we have used our assumption $[n_{F},Q]=0$; $n_{F}$ is Hermitian 
and positive definite so that 
$(n_{F})^{\pm 1/2}$ is well defined \footnote{The operator $(n_{F})^{- 1/2}$ is
well defined only at finite temperatures. However, one easily sees that
the spectra of $X$ and $\tilde{X}$ coincide even at zero temperature
by taking the corresponding limit.}).

The reality of the spectrum follows from the hermiticity of $\tilde{X}$.
The constraint on the eigenvalues follows from the observation that
both $\tilde{X}$ and $1-\tilde{X}$ are represented as sums of two
positive operators (see Eq.~(\ref{tX}) and the one obtained from it
by replacing $Q\to 1-Q$).

The spectral properties of the operator $X$ immediately imply
a constraint on the positions of zeros of the generating
function $\chi(\lambda)$ analytically continued to the
complex plane of the variable $u=e^{i\lambda}$, 
\begin{equation}
	\chi(u) = \det\left(\left[1+(u-1)X\right]u^{- Q}\right).
 \label{XuFCS}
\end{equation}
The zeros of $\chi(u)$ can be $0$, $\infty$  (from the factor $u^{-Q}$) 
or coincide with $1-X_{n}^{-1}$, where $X_{n}$ are the eigenvalues
of the operator $X$. From the spectral properties of $X$,
it then follows that
\textit{zeros of $\chi(u)$ are confined to the negative real axis 
of the complex $u$-plane}. This conclusion has been previously reached
in Ref.~\onlinecite{2008-AbanovIvanov} under the assumptions of zero 
temperature
and instant scattering. Generalizing it to arbitrary temperatures and
to arbitrary evolution operators (under the assumptions formulated
in Section~\ref{sec:detFCS}) constitutes the main result of our present paper. 

%%%%%%%%%%%%%%%%%%%%%%%%%%%%%%
\subsection{Generalized binomial statistics and effective transparencies}

The expression (\ref{XuFCS}) may be further interpreted in terms of
decomposing the charge transfer into a superposition of independent 
tunneling events, with the eigenvalues of $X$ giving
the probabilities of those tunneling processes. To make
this interpretation more explicit, we get rid of the last factor
$\exp[-i\lambda Q]$ in (\ref{XuFCS}) by
\footnote{We use the fact that $\tr\log(AB)=\tr\log A +\tr\log B$ if both 
terms are finite, i.e., if the operators $\log A$ and $\log B$ 
are of trace class.}
\begin{eqnarray}
\chi(\lambda) &=& \det\left( \left[1+(e^{-i\lambda}-1)X\right]
e^{-i\lambda X} e^{i\lambda X}e^{-i\lambda Q}\right)
 \nonumber \\
&=& e^{i\lambda\, \tr (X-Q)}
\det \left(\left[1+(e^{i\lambda}-1)X\right]e^{-i\lambda X}\right)
 \label{genbinom} \\
&=& e^{i\lambda \langle q \rangle}
\prod_n e^{-i\lambda X_n} [1+(e^{i\lambda}-1)X_n]\, ,
 \nonumber 
\end{eqnarray}
where in the last line the product is taken over the eigenvalues of the
operator $X$, and
\begin{equation}
	\langle q \rangle=
-i\partial_{\lambda}\ln\chi(\lambda)\Big|_{\lambda=0} =  
\tr (X-Q) = 
\tr\left[n_{F}(U^{\dagger}QU-Q)\right]
\end{equation}
is the average total transferred charge \cite{2007-AvronBachmannGrafKlich}.

Each factor  $[1+(e^{i\lambda}-1)X_n]$ in (\ref{genbinom})
corresponds to an elementary single-electron event with the
transmission probability $X_n$ and reflection probability $1-X_n$.
The other exponential factors in (\ref{genbinom}) correspond
to a deterministic ``background'' charge transfer, which does
not produce any noise.

Such a statistics given by a superposition of individual transmission
events with different transparencies was dubbed 
\textit{generalized binomial statistics} in Ref.~\onlinecite{2008-Hassler}.
Note that the effective transparencies are given by the spectrum of the 
\textit{effective-transparency operator} $X$ and depend in a complicated 
way on the time dynamics of the junction and on the
initial thermodynamic state.\footnote{Note that our results do not contradict
those of Refs.~\onlinecite{2008-VanevicNazarovBelzig,2007-VanevicNazarovBelzig}
where charge transport at zero temperature in the bias-voltage setup
has been factorized into one- and two-particle processes. Indeed, the
two-particle (``bidirectional'') processes in those works can be further
factorized into two symmetric sigle-particle transfers, in agreement with
our result.}

%%%%%%%%%%%%%%%%%%%%%%%%%%%%%%
\subsection{Spectrum of the effective-transparency operator}

Using the result derived above, we may now describe the full counting
statistics of non-interacting fermions by the distribution of effective
transparencies (the density of eigenvalues of the operator $X$), instead of the 
probabilities $P_q$.
Following Ref.~\onlinecite{2008-KlichLevitov}, we define the effective-transparency 
density as
\begin{equation}
	\mu(p) = \tr \delta(p-X)\, .
\end{equation}
Since the spectrum of $X$ defines the density of zeros of $\chi(u)$ on the negative
real semi-axis of $u$-plane, the density of states $\mu(p)$ may be related to the jump 
of the derivative of $\chi(\lambda)$ at the corresponding point. A simple
calculation gives \cite{2008-KlichLevitov} 
\begin{equation}
\mu(p)=
-\frac{1}{2\pi p(1-p)} \partial_{\lambda}\ln\chi(\lambda)\Big|^{u+i0}_{u-i0}
= \frac{1}{\pi}\I \partial_{p}\ln\chi(p-i0)\, ,
\label{jump-in-derivative}
\end{equation}
where 
\begin{equation}
p=\frac{1}{1-e^{i\lambda}}=\frac{1}{1-u} \, .
\end{equation}

Conversely, knowing the spectral density $\mu(p)$, 
one can reconstruct the cumulant generating function $\ln\chi(\lambda)$
(up to an overall deterministic charge transfer) by using (\ref{XlambdaFCS}):
\begin{equation}
\ln\chi(\lambda)  = -i\lambda \Tr Q 
+ \int_{0}^{1}dp\,\mu(p)\ln\left[1+(e^{i\lambda}-1)p\right] \, .
\label{CGFp}
\end{equation}

As for the structure of the spectrum of effective transparencies, 
there are two alternatives:
the spectrum may be either discrete or continuous.
An analysis of available examples gives the following possibilities:
\begin{enumerate}
\item
In a finite-dimensional problem (e.g., a finite number of incident
wave packets \cite{2008-Hassler}), the matrix is finite-dimensional,
and the spectrum of effective transparencies is obviously finite and 
discrete.
\item
At zero temperature, in the instant-scattering approximation,
for a continuous and periodic time dependence of the
scattering matrix \cite{2008-AbanovIvanov},
the spectrum is discrete, with possible accumulation points at $p=0$ and
$p=1$.
\item
At zero temperature, in the instant-scattering approximation,
for a discontinuous time dependence of the scattering matrix, the
spectrum typically becomes continuous \cite{2008-KlichLevitov}.
\item
From the above two examples, one can deduce that
at zero temperature, in the instant scattering approximation,
for a continuous non-periodic time dependence (a finite-time
pumping pulse), the spectrum is discrete, if the scattering matrix
returns to its initial value, and continuous otherwise. This
can be shown by a linear fractional mapping of the time axis on a circle \cite{1997-IvanovLeeLevitov},
thus reducing the non-periodic problem to a periodic one, with
either continuous or discontinuous time dependence of the scattering
matrix.
\item
For finite temperatures or energy-dependent scattering, the
spectrum of $X$ is typically continuous. This can be seen in
the examples presented in Section~\ref{sec:FCSnonint}. The physical
reason for a continuous spectrum is that particles incident at
different energies have different effective transparencies,
and thus integration over energies smears the spectrum.
\end{enumerate}
If the spectrum $\mu(p)$ is continuous, the singularities
form a branch cut of the function $\chi(u)$ along the negative real axis
(or along a part of it). At this branch cut, the function $\chi(u)$ 
is discontinuous, with the jump obeying (\ref{jump-in-derivative}).
A more detailed analysis of the spectrum of the effective transparencies
will be presented in a subsequent publication \cite{2009-AbanovIvanov-gap}.

Note that in the case of zero temperature and adiabatic pumping, the
spectrum of effective transparencies may alternatively be found from
a \textit{different} operator labeled $n_F^z$ in Ref.~\onlinecite{2008-AbanovIvanov}
and $M$ in Ref.~\onlinecite{2008-KlichLevitov}. It is worth emphasizing that
although the two operators are different, at zero temperature and
in the instant-scattering approximation their spectra coincide,
as shown in Appendix~\ref{app:adiabatic-nfz}.

%%%%%%%%%%%%%%%
%%%%%%%%%%%%%%%
\section{Examples: non-interacting fermions}
\label{sec:FCSnonint}
%%%%%%%%%%%%%%
%%%%%%%%%%%%%%

In this section, we illustrate our discussion with several known examples.

%%%%%%%%%%%%%
\subsection{$T=0$. Optimal (Lorentzian) pulses}

An instructive example where the full counting statistics 
is exactly computable is the case of several Lorentzian voltage
pulses of one-quantum intensity each: $\int V(t)\, dt = \pm 2\pi\hbar$.
In this case, as shown in Ref.~\onlinecite{1997-IvanovLeeLevitov}, the
full counting statistics involves only a finite number of transferred
electrons. For the effective-transparency spectrum, this
implies a finite discrete spectrum with the number of levels
equal to the number of pulses \cite{1997-IvanovLeeLevitov,2008-Sherkunov}. In the special
case when all pulses are of the same polarity (this also includes the
setup of a constant voltage \cite{1993-LevitovLesovik} as a limiting case), 
all the effective transparencies are degenerate and equal to the transparency
of the junction $g$:
\begin{equation}
\mu(p) = W\delta(p-g)\, ,
\end{equation}
where $W=(2\pi\hbar)^{-1} \int V(t)\, dt $ is the number of pulses.

%%%%%%%%%%%%%
\subsection{$T=0$. Periodic opening of a quantum point contact}

\begin{figure}
\includegraphics[width=8cm]{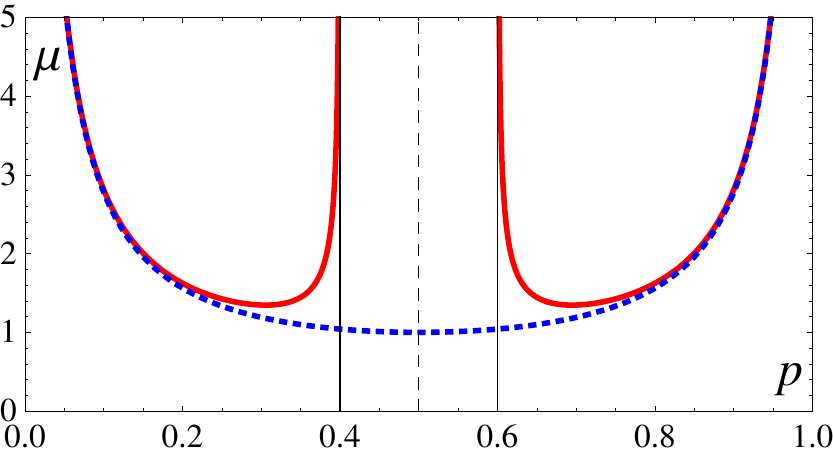}
\caption{The effective-transparency distribution (\ref{muinvpT0})
for $g=1$ (blue dashed line) and $g=0.96$ (red solid line), plotted 
in the units of $4 G/\pi^{2}$. 
For $g=0.96$, there is a spectral gap between $p=0.4$ and $p=0.6$.}
\label{fig:gap}
\end{figure}

An interesting exactly solvable example with a \textit{continuous} spectrum
of effective transparencies has been considered in
Ref.~\onlinecite{2008-KlichLevitov}: a periodic train of  
openings and closings of the contact. In this case, the distribution
of effective transparencies is (Fig.~\ref{fig:gap})
\begin{equation}
\mu(p) = \frac{G}{\pi^{2}} \frac{\sqrt{g}}{p(1-p)} \,
\R \frac{|1-2p|}{\sqrt{(1-2p)^{2}-(1-g)}},
 \label{muinvpT0}
\end{equation}
where the overall intensity [per opening/closing cycle, 
as in (\ref{chi-extensive})] is
\begin{equation}
G = 2\ln \frac{\sin\pi w}{\pi w_{0}}.
 \label{GT0}
\end{equation}
Here $w$ is the fraction of a period during which the contact 
is open (with the transparency $g$) and $w_{0}$ is a regularization given, 
essentially, by the fraction of a period taken by the switching time. 

One can see that if the contact is not fully open ($g<1$), 
then there is a spectral gap around $\mu(p)=1/2$.
We will comment on this feature in a future 
publication\cite{2009-AbanovIvanov-gap}.

%%%%%%%%%%%%%%%%
%%%%%%%%%%%%%%%%

%%%%%%%%%%%%%%%%%
\subsection{$T\neq 0$. Zero bias voltage}

Remarkably, the same formula (\ref{muinvpT0}) gives the distribution
of effective transparencies in the case of a finite temperature $T$
(and zero bias voltage) \cite{1996-LevitovLeeLesovik,2008-KlichLevitov}.
The overall intensity is then given by
\begin{equation}
	G = \frac{\pi t T}{\hbar}\, ,
 \label{GT}
\end{equation}
and the observation time $t$ is assumed to be much longer than $\hbar/T$,
so that only the extensive part (\ref{chi-extensive}) is retained.

%%%%%%%%%%%%%%%%%%%%%%%%%%%%%%
%%%%%%%%%%%%%%%%%%%%%%%%%%%%%%
\section{Examples: interacting fermions}
 \label{sec:FCSint}
%%%%%%%%%%%%%%%%%%%%%%%%%%%%%%
%%%%%%%%%%%%%%%%%%%%%%%%%%%%%%

The derivation of Section \ref{sec:FCSzeros}  shows that the absence of interactions
between fermions leads to a generalized binomial statistics of the charge transfer 
between two conductors. In this section, we show that if the fermions interact,
the zeros of the generating function $\chi(u)$ may shift to the complex plane.

Before presenting examples of full counting statistics for interacting
fermions, let us use a two-particle example to gain some intuition about
the positions of the roots of $\chi(u)$.
Consider a system of two particles which
propagate from the left to the right lead of the contact. The full counting
statistics is then given by a quadratic polynomial 
$\chi(u) =P_{0}+P_{1}u+P_{2}u^{2}$ with the probabilities $P_{0,1,2}$ 
obeying $P_{0}+P_{1}+P_{2}=1$. The roots of the polynomial become non-real if
\begin{equation}
P_{1}^{2} < 4 P_{0}P_{2}\, ,  \qquad \mbox{or, equivalently,}\qquad 
\sqrt{P_{0}}+\sqrt{P_{2}} > 1\, ,
 \label{complcond}
\end{equation}
i.e., if the probability $P_1$ of separating the two particles becomes too small.
If one considers a strong attraction between particles, so that they become inseparable,
then $P_1=0$, and the roots of $\chi(u)$ become purely imaginary. One thus sees that
in this example (i) the non-reality of roots may be loosely related to an attraction
between particles; (ii) a certain threshold of interaction may be needed to shift
the roots of $\chi(u)$ from the real axis. 

We now turn to several more complicated examples of interacting systems where the 
full counting statistics may be exactly computed.

%%%%%%%%%%%%%%%%%
\subsection{Charge transfer in a normal metal -- superconductor point contact}

A more general example of an electronic system with attraction leading to charge transfer
quantized in pairs of electrons (and hence to a shift of the zeros of $\chi(u)$ from the
real axis) is considered in Ref.~\onlinecite{1994-MuzKhm}: a contact between a 
normal metal and a superconductor. In the low-temperature limit and at low bias voltage $V$
(so that $T\ll eV\ll \Delta$), single-particle tunneling is suppressed, and the full 
counting statistics is given by
\begin{equation}
\chi(u) = (1-g_{A} +g_{A}u^{2})^{W}\, .
 \label{chiNS}
\end{equation}
Here $g_{A}$ is the probability of Andreev reflection and $W=eVt/(2\pi\hbar)$ is the effective
number of transfer attempts during the observation time $t$. Similarly to our 
two-particle example above, the zeros of $\chi(u)$ are in this case purely imaginary.

%%%%%%%%%%%%%%%%%
\subsection{Two particles scattering on a resonant quantum dot}

Another example of an interacting two-particle scattering problem has been studied in
Ref.~\onlinecite{2008-LebedevLesovikBlatter}. In that work, one considers a quantum dot characterized
by its resonances and by a Coulomb interaction between the particles. In this model,
the scattering matrix is exactly computed and then used to find the full counting
statistics in a two-particle example. For the example considered in that work
(a singlet two-particle wave packet in the shape of an exponentially truncated plane wave),
the probabilities to transfer one and two electrons are given by (we
refer the reader to Ref.~\onlinecite{2008-LebedevLesovikBlatter} for details of the model
and the derivation)
\begin{eqnarray}
P_{1} &=& \frac{2\beta}{(1+3\beta)(1+\beta)^{2}}\left[2+3\beta
-\frac{1}{\delta^{2}+1}\right] ,
\\
 P_{2} &=&  \frac{1}{(1+3\beta)(1+\beta)^{2}} \left[
1+
\frac{s\, \beta\, [4\delta^{2}+3(3+\beta)]}
{2(1+\beta) (\delta^{2}+1)(\delta^{2}+s^{2})}\right]\, ,
\nonumber
\end{eqnarray}
where  $\beta$ is the dimensionless dwelling time on the dot characterizing the resonance, $\delta$
characterizes the strength of the Coulomb interaction (it is related to the parameter $\alpha$
introduced in Ref.~\onlinecite{2008-LebedevLesovikBlatter} as  $\delta= \alpha\beta/(1+\beta)$), and
$s=(3+\beta)/[2(1+\beta)]$.

\begin{figure}
\includegraphics[width=8cm]{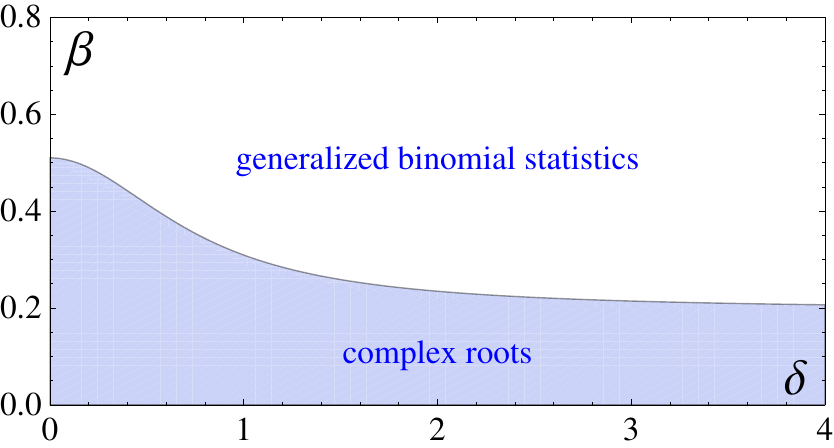}
\caption{The shaded area shows the region in the parameter space of $\delta$ and $\beta$,
where the roots of $\chi(u)$ are complex and the statistics is \textit{not} generalized 
binomial.}
\label{fig:pair-tunneling}
\end{figure}

Using the criterion (\ref{complcond}), we can find the range of parameters 
$\delta$ and $\beta$ where the statistics is generalized binomial 
(see Fig.~\ref{fig:pair-tunneling}). Depending on the dwelling time, an
interaction may either shift the roots of $\chi(\lambda)$ from 
the real axis or not. 

Although this is a two-particle example, a qualitatively similar behavior 
of roots can also be found in multiparticle systems.  
For example, in Refs.~\onlinecite{2005-KomnikGogolin,2007-SGK},
full counting statistics has been calculated for 
charge and spin transport through a Kondo dot. In that problem, 
one can have both regimes with roots real and complex,
depending on the parameters.

%%%%%%%%%%%%%%%%
%%%%%%%%%%%%%%%%

%%%%%%%%%%%%%%%%%
\subsection{Charge pumping through a single-electron transistor}

An interesting example with interaction where the statistics is
nevertheless generalized binomial is studied in Ref.~\onlinecite{2001-AndreevMishchenko}.
The authors of that work consider a charge pump based on a nearly open quantum dot 
in the Coulomb blockade regime. In  spite of the interaction on the dot, 
the problem is mapped onto an effective quadratic fermionic Hamiltonian (including a 
Majorana fermion representing the resonant level at the dot). 
Therefore, our formal conclusion about the
generalized binomial statistics remains valid for this interacting system as well.

Indeed, one can easily verify that the generating function (per period) found in 
Ref.~\onlinecite{2001-AndreevMishchenko},
\begin{equation}
\ln \chi(\lambda) =-i\lambda +  \frac{1}{2\hbar\omega}
\int d\epsilon\, \ln \left( 1 - \frac{\epsilon^2}{\epsilon^2+\Gamma^2}
\left[\frac{n_-(1-n_+)}{p} + \frac{n_+(1-n_-)}{1-p}\right]\right)\, ,
 \label{AMlimit}
\end{equation}
contains only real roots $p$ (at any temperature). Here $\omega$ is the pumping frequency,
$n_{\pm}(\epsilon) = (e^{(\epsilon\pm \hbar\omega)/T}+1)^{-1}$ are the Fermi occupation 
numbers at the temperature $T$, $\Gamma$ is an energy scale characterizing the strength of 
the Coulomb interaction in the dot, and $p=(1-e^{i\lambda})^{-1}$, as usual.

\begin{figure}
\includegraphics[width=8cm]{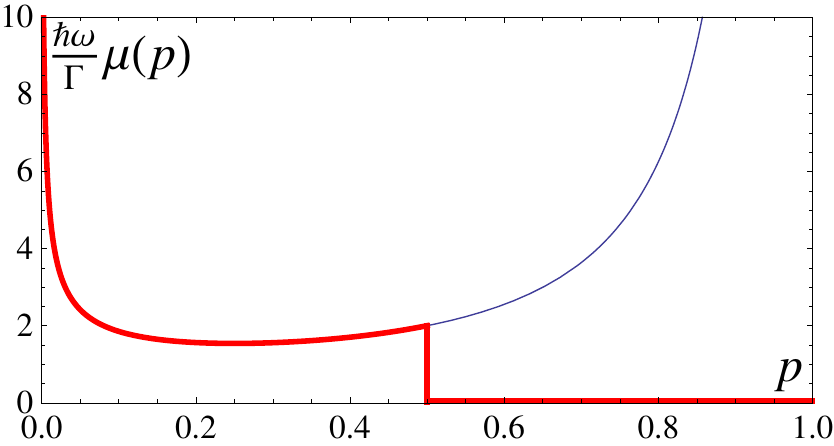}
\caption{The effective-transparency distribution (\ref{AMmupT0}), in the units 
of $\Gamma/\hbar\omega$ (thick red line). $\mu(p)$ vanishes abruptly at 
$p=(\hbar\omega)^{2}/((\hbar\omega)^{2}+\Gamma^{2})$ (shown for $\hbar\omega=\Gamma$).  }
\label{fig:Coulomb}
\end{figure}

In the zero-temperature limit, the effective-transparency distribution $\mu(p)$ can
be easily calculated analytically:
\begin{equation}
\mu(p) = \begin{cases}
\frac{\Gamma}{2\hbar\omega}\, p^{-1/2}(1-p)^{-3/2} & 
\text{for $p<\frac{(\hbar\omega)^{2}}{(\hbar\omega)^{2}+\Gamma^{2}}$ 
\, ,} \\
0 & \text{for $p>\frac{(\hbar\omega)^{2}}{(\hbar\omega)^{2}+\Gamma^{2}}$ \, .}
\end{cases}
 \label{AMmupT0}
\end{equation}
The spectrum is continuous with a gap at
$\frac{(\hbar\omega)^{2}}{(\hbar\omega)^{2}+\Gamma^{2}}<p<1$,
as shown in Fig.~\ref{fig:Coulomb}
(one can check that the gap survives at finite temperatures, but changes its
magnitude and shifts towards the center of the interval).

%%%%%%%%%%%%%%%%
%%%%%%%%%%%%%%%%

%%%%%%%%%%%%%%%%%%%%%%%%%%%%%%
%%%%%%%%%%%%%%%%%%%%%%%%%%%%%%
\section{Conclusion}
%%%%%%%%%%%%%%%%%%%%%%%%%%%%%%
%%%%%%%%%%%%%%%%%%%%%%%%%%%%%%
\label{sec:Conclusion}

To summarize, we have shown that the charge-transfer statistics for
a system of non-interacting fermions is generalized binomial: it factorizes
into individual single-particle tunneling events. This property holds
under very general conditions: at arbitrary temperature and for
an arbitrary time- and energy-dependent scatterer. Note that the
converse is not true: an interacting system may either obey or disobey
this factorization property, as we have seen in examples.

The factorization property may be formulated in terms of the positions
of zeros of the generating function $\chi(u)$ (or, equivalently, of the
singularities of $\ln\chi(u)$). The factorization implies that those
singularities are confined to the negative real axis. In this case one can 
fully characterize full counting statistics by the distribution of effective
transparencies $\mu(p)$. All charge-transfer properties 
(moments and cumulants of the transferred charge, entanglement between the
leads \cite{2008-KlichLevitov}, etc.) can be expressed in terms of $\mu(p)$. 

The absence of factorization corresponds to a shift of singularities from the real axis. 
In this case, $\chi(u)$ can still be characterized by its zeros albeit in the complex $u$-plane.
This description is somewhat similar to the approach of Lee and Yang in statistical physics, 
where a partition function is characterized by the zeros of its analytic extension to complex values 
of physical parameters such as magnetic field or chemical potential\cite{1952-LeeYang}. 
In statistical physics, this approach turned out to be very useful for understanding possible 
phase transitions: a phase transition is determined by the location of zeros of the partition function
relative to the real (physical) axis of parameters \cite{1952-LeeYang}. From this
perspective, there remain interesting questions of characterizing interactions in the 
scattering problem by the positions of singularities of $\ln\chi(u)$ and 
of understanding possible physical implications of their shift away from the real axis.

As an interesting extension of our results, one can consider
constraints on full counting statistics for non-interacting electrons
in a multi-terminal setup. In this case, a certain ``reality condition''
on the zeros of the generating function is possible \cite{2009-Kambly}.
However, the factorization property does not hold, being specific to
the two-terminal case.

Another possible generalization of our construction is to use the formalism
developed in Sections \ref{sec:detFCS} and \ref{sec:FCSzeros} for other types
of problems with quadratic Hamiltonians, not necessarily involving physical
charge transfer. As an example of such an application, one can study the statistics 
of staggered magnetization of a XY spin chain (mappable to free fermions by a Jordan-Wigner
transformation). See also Ref.~\onlinecite{2006-Klich} for applications of those ideas to 
analyzing the entropy in such systems.

%%%%%%%%%%%%%%%%%%%%%%%%%%%%%%%%%%%%%%%%%%%%%%%%%%%%%%%%%%%%%
\section{Acknowledgments}
%%%%%%%%%%%%%%%%%%%%%%%%%%%%%%%%%%%%%%%%%%%%%%%%%%%%%%%%%%%%%

We have  benefited from discussions
with S.~Bieri, D.~Kambly, G.~Lesovik, L.~Levitov, Yu.~Makhlin, and E.~Sukhorukov. 
A.G.A. is grateful to ITP, EPFL for hospitality in Summer 2008.  
The work of A.G.A. was supported by the NSF under the grant DMR-0348358.

%%%%%%%%%%%%%%%%%%%%%%%%%%%%%%%%%%%%%%%%
\appendix

%%%%%%%%%%%%%%%%%%%%%%%%%%%%%%%%%%%%%%%%
\section{Fermionic bilinears and trace identities}
\label{app:fbil}
%%%%%%%%%%%%%%%%%%%%%%%%%%%%%%%%%%%%%%%%

For reader's convenience, we present in this appendix some helpful identities
on the exponentials of fermionic bilinears. Those identities are then used
to derive the determinant formula (\ref{spFCS}). A more detailed account including proofs
may be found in Ref.~\onlinecite{2002-Klich}.

For any single-particle observable $A$, the corresponding multi-particle operator acting in the 
Fock space is given by the \textit{fermionic bilinear}
\begin{equation}
	\hat{A} = \psi^{\dagger}A \psi \, .
\end{equation}
This construction provides a representation of the $gl(n)$ algebra in the 
$2^{n}$-dimensional Fock space ($n$ denotes here the dimension of the single-particle Hilbert
space). Indeed, one easily checks that commutators of fermionic bilinears are given by
\begin{equation}
	[\hat{A},\hat{B}] = \widehat{\left[A,B\right]}\, .
\end{equation}
Exponentiating these relations, one obtains a representation of the group $GL(n)$:\footnote{One can prove 
(\ref{groupP}) by using Baker--Campbell--Hausdorff formula. See, e.g., 
Yu.~A.~Bakhturin, ``Campbell--Hausdorff formula'', in M.~Hazewinkel, 
Encyclopaedia of Mathematics, Kluwer Academic Publishers, 2002.}
\begin{equation}
	e^{\hat{A}} e^{\hat{B}} = e^{\hat{C}}, \qquad \mbox{if}\qquad e^{A}e^{B}=e^{C}.
 \label{groupP}
\end{equation}

One can use this representation to show that the evolution operator for a system
of noninteracting
fermions is an exponential of a fermionic bilinear. Indeed, the multi-particle
evolution operator is given by the time-ordered exponential
\begin{equation}
	\hat{U} = {\rm Texp}\left[ -i \int \hat{H}(t)\, dt \right],
\end{equation}
where the time-dependent Hamiltonian $\hat{H}$ is a fermionic bilinear. 
This time-ordered exponential can be represented as a chronologically 
ordered product of (infinitesimal) exponents of fermionic bilinears and, by virtue of
(\ref{groupP}) may be rewritten as a single exponential
\begin{equation}
\hat{U}=\exp[\psi^\dagger (\ln U) \psi]\, ,
\end{equation}
where
\begin{equation}
U = {\rm Texp}\left[ -i \int H(t)\, dt \right]
\end{equation}
is the \textit{single-particle} evolution operator.

For traces of such exponentials one has a simple formula \cite{2002-Klich}:
\begin{equation}
	\Tr e^{\hat{A}} = \det(1+e^{A})\, ,
 \label{1Mid}
\end{equation}
where the trace is taken in the Fock space, and the determinant is 
in the single-particle space.
For traces of products of exponentials of fermionic bilinears, one finds,
using (\ref{groupP}) and (\ref{1Mid}):
\begin{equation}
	\Tr\left(e^{\hat{A}_1} e^{\hat{A}_2} \dots e^{\hat{A}_k}\right) = 
  \det(1+e^{A_1}e^{A_2} \dots e^{A_k})\, .
 \label{2Mid}
\end{equation}

From this relation, a derivation of the Levitov--Lesovik determinant formula
easily follows: 
\begin{equation}
\frac{\Tr\left(\hat{\rho}_{0} \hat{U}^{\dagger}e^{i\lambda\hat{Q}}\hat{U}
e^{-i\lambda \hat{Q}}\right)}{\Tr\hat{\rho}_{0}} =
\frac{\det\left(1+e^{-h_{0}} U^{\dagger} e^{i\lambda Q}U e^{-i\lambda Q}
\right)}{\det\left(1+e^{-h_0} \right)} =
\det\left(1 +n_{F}\left(U^{\dagger} e^{i\lambda Q}U e^{-i\lambda Q}-1\right)\right)\, ,
 \label{Levderiv}
\end{equation}
where $n_F$ is defined by (\ref{nF}).

%%%%%%%%%%%%%%%%%%%%%%%%%%%%%%%%%%%%%%%%
\section{Adiabatic and instant-scattering approximations}
\label{app:adiabatic}
%%%%%%%%%%%%%%%%%%%%%%%%%%%%%%%%%%%%%%%%

The determinant formula (\ref{spFCS}) has been derived under the very general assumptions specified
in Section~\ref{sec:detFCS}. A physically relevant and technically simpler approximation is 
possible in the
situation when the two leads involve asymptotically free electrons, so that the scattering 
may be described
in terms of an instantaneous scattering matrix $S(t,\epsilon)$. This matrix relates the asymptotic states
for electrons at energy $\epsilon$ scattering on the instantaneous Hamiltonian $H(t)$. If the Hamiltonian
varies slowly on the scale of the ``scattering time''  $\hbar\, \partial \log S/ \partial \epsilon $, then
the evolution operator $U$ may be approximated by the Wigner transformation of $S(t,\epsilon)$ 
\cite{2003-MuzykantskyAdamov}. The corresponding \textit{adiabaticity condition} may be formulated as
\begin{equation}
\hbar\, \left\vert\frac{\partial S^{-1}}{\partial t}\frac{\partial S}{\partial \epsilon} \right\vert \ll 1 \, .
 \label{adiabcrit}
\end{equation}
At zero temperature, this condition allows to neglect the energy dependence of $S(t,\epsilon)$, and
to replace the scattering operator $U$ in (\ref{spFCS}) by the instantaneous scattering matrix $S(t)$
taken at the Fermi energy $\epsilon$ (since only the states in the vicinity of the Fermi
energy contribute to the determinant) \cite{1993-IvanovLevitov,1997-IvanovLeeLevitov,2003-MuzykantskyAdamov}.

At finite temperature, to neglect the energy dependence of the scattering matrix, one also needs that
the energy dependence of $S$ is small on the scale determined by the temperature. In addition to the 
adiabaticity condition (\ref{adiabcrit}) we require:
\begin{equation}
	T \, \left\vert  S^{-1}\, \frac{\partial S}{\partial \epsilon} \right\vert \ll 1 \, .
 \label{instant-temp}
\end{equation}
The two conditions (\ref{adiabcrit}) and (\ref{instant-temp}) constitute the \textit{instant-scattering}
approximation (equivalent to the adiabatic approximation in the case of zero temperature). In this approximation,
the determinant formula becomes
\begin{equation}
\chi(\lambda) = \det\left(1+n_{F}(S^{\dagger}e^{i\lambda Q}Se^{-i\lambda Q}-1)\right)\, ,
 \label{spFCSadiab}
\end{equation}
where $S(t)$ is a local in time unitary scattering matrix.

The determinant (\ref{spFCSadiab}), in turn, requires a regularization at negative energies, since
the operator under the determinant tends at infinite negative energies to
$S^{\dagger}e^{i\lambda Q}S e^{-i\lambda Q}$, a non-unity matrix. This regularization may be performed
either by taking into account the finiteness of the negative-energy spectrum or
by re-introducing a weak energy dependence of $S(t,\epsilon)$ so that 
$\lim_{\epsilon\to -\infty} S(t,\epsilon)= 1$. In the latter case, the easiest way
to implement the regularization is to rewrite (\ref{spFCS}) as \cite{2007-AvronBachmannGrafKlich}
\begin{equation}
\chi(\lambda) = \det\Big(\left[1+n_{F}(U^{\dagger}e^{i\lambda Q} U e^{-i\lambda Q}-1)\right] \cdot
e^{i\lambda n_F Q}U^{\dagger} e^{-i\lambda n_F Q}U \Big) \, .
 \label{adiab-regularized}
\end{equation}
This manipulation is obviously admissible for the ``full'' evolution operator $U$ corresponding
to the regularized scattering matrix with $\lim_{\epsilon\to -\infty} S(t,\epsilon)= 1$. However,
in the new expression (\ref{adiab-regularized}), one can simply replace $U$ by a local in time 
scattering matrix $S(t)$ to obtain a fully convergent expression in the instant-scattering
limit.

The regularized formula (\ref{adiab-regularized}) proposed in Ref.~\onlinecite{2007-AvronBachmannGrafKlich}
is equivalent to the previously existing regularization prescriptions.
Our discussion in this paper does not rely on this regularization, since we always consider
the general case of an arbitrary non-interacting evolution operator and therefore may
assume that the negative-energy asymptotics is suitably regularized.

%%%%%%%%%%%%%
\section{Spectrum of the effective transparencies in the adiabatic limit at zero temperature}
\label{app:adiabatic-nfz}
%%%%%%%%%%%%%

Generalized binomial statistics for non-interacting fermions
has already been proven in our earlier publication under
the assumption of zero temperature and instant scattering \cite{2008-AbanovIvanov}. 
Of course, that result now follows from the more general argument of the present work. 
However, we find it instructive to demonstrate explicitly the equivalence of the two results
under those more restrictive assumptions.

It has been shown in Ref.~\onlinecite{2008-AbanovIvanov} that, at zero temperature and in the
instant-scattering limit, the effective transparencies of elementary single-particle
events are given by the spectrum of the operator
\begin{equation}
n_F^z = Q S n_F S^\dagger Q
\label{nfz}
\end{equation}
(the same operator was denoted $M$ in Ref.~\onlinecite{2008-KlichLevitov}).
In this Appendix,
we explicitly show that the spectra of $n_F^z$ and of $X$ coincide, except for the
eigenvalues zero and one.

Indeed, at zero temperature $n_F$ is a projector operator, and therefore the ``hermitized''
operator $\tilde{X}$ given by (\ref{tX}), whose spectrum coincides with the spectrum of $X$,
is block-diagonal. The block at positive energies equals $Q$ and only gives eigenvalues zero
and one. The block at negative energies can be written (at zero temperature) as
\begin{equation}
\tilde{X}_- = n_F S^\dagger Q S n_F = R^\dagger R \, , \qquad \mbox{where} \qquad
R = Q S n_F\, 
\end{equation}
(we have also used that $Q$ is a projector). On the other hand, one
can rewrite (\ref{nfz}) as
\begin{equation}
n_F^z = R R^\dagger\, ,
\label{nfz2}
\end{equation}
from where it becomes obvious that the spectra of $\tilde{X}_-$ and
$n_F^z$ coincide, except for zero eigenvalues (their eigenvectors can be
related to each other by the operators $R$ and $R^\dagger$). This completes the
proof of the coincidence of the spectra of $X$ and $n_F^z$ (modulo eigenvalues
zero and one, which do not contribute to the noise, but only to the overall 
deterministic charge transfer).

%%%%%%%%%%%%%%%%%%%%%%%%%%%%%%%%%%%%%%%%

%%%%%%%%%%%%%%%%%%%%%%%%%%%%%%%%

\end{document}